\documentclass[twocolumn,english]{revtex4}
\usepackage[T1]{fontenc}
\usepackage[latin9]{inputenc}
\usepackage{babel}
\usepackage{amsmath}
\usepackage{amssymb}
\usepackage{graphicx}
\usepackage{esint}
\usepackage[unicode=true,pdfusetitle,
 bookmarks=true,bookmarksnumbered=false,bookmarksopen=false,
 breaklinks=false,pdfborder={0 0 1},backref=section,colorlinks=false]
 {hyperref}
\usepackage{breakurl}

\makeatletter
\@ifundefined{textcolor}{}
{%
 \definecolor{BLACK}{gray}{0}
 \definecolor{WHITE}{gray}{1}
 \definecolor{RED}{rgb}{1,0,0}
 \definecolor{GREEN}{rgb}{0,1,0}
 \definecolor{BLUE}{rgb}{0,0,1}
 \definecolor{CYAN}{cmyk}{1,0,0,0}
 \definecolor{MAGENTA}{cmyk}{0,1,0,0}
 \definecolor{YELLOW}{cmyk}{0,0,1,0}
}

\usepackage{bbold}

\makeatother

\begin{document}

\title{Characterizing eigenstate thermalization via measures in the Fock
space of operators}

\author{Pavan Hosur}

\affiliation{Department of Physics, Stanford University, Stanford, CA 94305-4045,
USA}

\author{Xiao-Liang Qi}

\affiliation{Department of Physics, Stanford University, Stanford, CA 94305-4045,
USA}
\begin{abstract}
The eigenstate thermalization hypothesis (ETH) attempts to bridge
the gap between quantum mechanical and statistical mechanical descriptions
of isolated quantum systems. Here, we define unbiased measures for
how well the ETH works in various regimes, by mapping general interacting
quantum systems on regular lattices onto a single particle living
on a high-dimensional graph. By numerically analyzing deviations from
ETH behavior in the non-integrable Ising model, we propose a quantity
that we call the $n$\emph{-weight} to democratically characterize
the average deviations for all operators residing on a given number
of sites, irrespective of their spatial structure. It\emph{ }appears
to have a simple scaling form, that we conjecture to hold true for
all non-integrable systems. A closely related quantity, that we term
the $n$\emph{-distinguishability}, tells us how well two states can
be distinguished if only $n$-site operators are measured. Along the
way, we discover that complicated operators on average are worse than
simple ones at distinguishing between neighboring eigenstates, contrary
to the naive intuition created by the usual statements of the ETH
that few-body (many-body) operators acquire the same (different) expectation
values in nearby eigenstates at finite energy density. Finally, we
sketch heuristic arguments that the ETH originates from the limited
ability of simple operators to distinguish between quantum states
of a system, especially when the states are subject to constraints
such as roughly fixed energy with respect to a local Hamiltonian. 
\end{abstract}
\maketitle

\section{Introduction}

Statistical mechanics and quantum mechanics have been the cornerstones
of modern physics for nearly a century. Both formalisms have been
put to test in a wide variety of scenarios, and both have invariably
given descriptions that are accurate as well as consistent with each
other within their regimes of validity. Their starting points, however,
are fundamentally different. Quantum mechanics describes unitary evolution
of isolated systems via wave functions or pure states, whereas statistical
mechanics ascribes microcanonical ensembles -- which are inherently
mixed states -- to them. Why, then, do the two theories concur?

A crucial step towards answering this question was taken by Berry
\cite{Berry1977}, who conjectured that the eigenstates of quantum
Hamiltonians whose classical counterparts are chaotic behave as if
they were drawn randomly from a Gaussian distribution. In other words,
they resemble random superpositions of classical configurations at
the same energy density in the thermodynamic limit. Deutsch \cite{Deutsch1991}
argued that Berry's conjecture holds for integrable systems perturbed
away from integrability. The argument stems from the intuition that
integrability-breaking perturbations allow quantum systems to mimic
the central maxim of classical statistical mechanics. On the one hand,
classical ergodic systems explore all accessible microstates with
equal likelihood within the time scale of typical measurements and
hence, justify equating time and ensemble averages. On the other,
quantum systems can access all classical ``microstates'' \emph{at
once} from a single eigenstate, so no time or ensemble-averaging is
required and each eigenstate effectively resembles a microcanonical
ensemble. Srednicki \cite{Srednicki1994} then explicitly derived
statistical distributions for quantum particles assuming only Berry's
conjecture, and postulated the \emph{eigenstate thermalization hypothesis}
(ETH). The ETH states, 
\begin{quote}
In ergodic quantum systems, eigenstates at finite energy density give
rise to expectation values for ``simple operators'' that vary smoothly
with energy, with fluctuations that are exponentially suppressed in
the system size. The off-diagonal matrix elements of these operators
between nearby eigenstates also vanish exponentially. 
\end{quote}
Here, ``simple operators'' usually refers to operators that involve
a very small number of degrees of freedom compared to the system size,
although the precise definition is given very rarely \cite{Garrison2015}.

The ETH is only believed to hold for simple operators and states with
finite energy density in the thermodynamic limit. With increasing
complexity of operators, their ability to tell eigenstates with nearby
energy apart improves. The extreme case is that of a projection operator
$\rho_{n}=|n\rangle\langle n|$ onto a given eigenstate $|n\rangle$;
this typically has support on all sites of the system and it obviously
distinguishes $|n\rangle$ from all other eigenstates perfectly. Therefore,
a refined statement of the ETH demands the definition of quantitative
measures for how distinguishable two states are when a restricted
set of operators is measured. Moreover, we will see that choosing
this restricted set to be all operators of certain complexity (rather
than some arbitrarily chosen operators) permits the definition of
a ``basis-independent measure'' to systematically quantify how well
the ETH works and how distinguishable nearby states are as a function
of the system size, eigenstate energy and operator complexity.

In this work, we define two such measures: 
\begin{enumerate}
\item the $n$-weight of a Hermitian operator $A$, 
\begin{equation}
P_{n}(A)=\sum_{\ell}\left|\mbox{Tr}\left(AO_{\ell}^{(n)}\right)\right|^{2}
\end{equation}
where $O_{\ell}^{(n)}$ for all $\ell$ form a complete orthonormal
basis (with respect to the Hilbert-Schmidt norm) of operators that
have support on $n$ sites, and 
\item the $n$-distinguishability of two operators $A_{1}$ and $A_{2}$,
\begin{equation}
\theta_{n}(A_{1},A_{2})=\cos^{-1}\frac{P_{n}(A_{1})+P_{n}(A_{2})-P_{n}(A_{1}-A_{2})}{2\sqrt{P_{n}(A_{1})P_{n}(A_{2})}}
\end{equation}

\end{enumerate}
These measures are defined for a lattice system with the Hilbert space
$\mathbb{H}=\prod_{i}^{\otimes}\mathbb{H}_{i}$, $\mathbb{H}_{i}$
being the Hilbert space of the $i^{th}$ site. The intuition behind
these quantities is an alternative view of quantum many-body dynamics.
As will be described in Sec. \ref{sec:fock-space}, Heisenberg evolution
of a many-body operator can be mapped to a quantum mechanical problem
of a single particle hopping on a high-dimensional graph. Each site
of the graph represents a basis operator, e.g., a direct product of
Pauli operators if each lattice site hosts a single qubit. The "particle
wavefunction" is determined by the overlap of operator $A$ with
the basis operator. The graph has a layered structure with all $n$-site
operators $O_{\ell}^{(n)}$ in the $n$-th layer. In this picture,
the $n$-weight $P_{n}(A)$ is simply proportional to the probability
of the particle being in the $n$-th layer of the graph. The $n$-distinguishability
of two operators $A_{1}$ and $A_{2}$ is the angle between the projections
of the corresponding vectors in the Hilbert space of operators onto
the subspace of the graph sites that constitute the $n^{th}$ layer.
A large angle $\theta_{n}$ means $A_{1}$ and $A_{2}$ are easier
to distinguish with $n$-site operators.

Most previous studies have numerically analyzed the expectation values
or off-diagonal matrix elements of a select set of simple operators
dwelling in contiguous regions of space either at long times \cite{Rigol2008,Rigol2009a,Rigol2012,Rigol2014,Khlebnikov2014,Sorg2014,Marcuzzi2013}
or in different eigenstates \cite{Kim2014,Khemani2014,Alba2015,Beugeling2015,Beugeling2014,Ikeda2011}.
Infinite temperature studies are the only exceptions that we are aware
of that include non-contiguous regions of space \cite{Lubkin1978,Page1993,Foong1994,Sanchez-Ruiz1995,Sen1996}.
In contrast, our measures treat all operators with support on a given
number of sites on equal footing, irrespective of whether the sites
are adjacent or spatially separated. If $A$ is chosen to be a density
operator, then the $n$-weight turns out to be closely related to the
second Renyi entropy of regions with size $n$ and smaller. However,
the Renyi entropy of an $n$-site region has contributions from fewer
site operators too, while the $n$-weight is a direct measure of the
amplitude of $n$-site correlation functions in this density operator.

Using the two measures we define, we study two related quantities
that characterize the difference between two neighboring eigenstates,
with density operators $\rho_{1}$ and $\rho_{2}$. The first quantity
is the $n$-weight of the difference, $P_{n}\left(\rho_{1}-\rho_{2}\right)$.
By performing numerical exact diagonalization on a simple non-integrable
model, an Ising model with parallel and transverse fields, we discover
that $P_{n}\left(\rho_{1}-\rho_{2}\right)$ attains a simple scaling
form as a function of $n$ and the density of states in the part of
the spectrum from which the two states are drawn. We conjecture that
a similar form holds in all non-integrable systems. Our result on
$P_{n}\left(\rho_{1}-\rho_{2}\right)$ also clarifies a rather peculiar
feature of quantum ergodic systems. According to the ETH, one would
naively think that a complicated (high $n$) operator has a better
chance of distinguishing two neighboring eigenstates than a simple
operator. We discover that surprisingly, the opposite is true for
typical operators: simple operators \emph{on average} are better\emph{
}at distinguishing between nearby eigenstates than complicated operators
are. In fact, there is a critical size of operators, which only depends
on the system size and the Hilbert space dimension at each site, at
which measuring a random operator reveals no information about the
system. This critical operator size occurs well beyond half of the
system size, and was consequently missed by previous studies that
focused on simple operators.

The second quantity we study is the $n$-distinguishability for the
same pair of eigenstates, $\theta_{n}(\rho_{1},\rho_{2})$, which
helps to reconcile the intuition and the numerical observation mentioned
above. This angle is found to be small when the vectors are projected
onto a low-$n$ subspace, determined by simple operators, but to saturate
to $\pi/2$ for more complex operators. Both these behaviors are found to be strikingly
different in integrable systems. The large $n$-distinguishability
suggests that although most large operators are poor at distinguishing
between neighboring eigenstates, there are a few fine-tuned ones that
are proficient at it. Thus, the optimal $n$-site operator can differentiate
between the two states better if $n$ is large, but a randomly chosen
one works better for smaller $n$. The combination of the two measures
$P_{n}$ and $\theta_{n}$ thus provides a more systematic and accurate
understanding to the nature of thermalization.

The remainder of the paper is organized as following. In Sec. \ref{sec:fock-space}
we describe the Fock space of operators and set the general framework
of our approach. The two quantities $n$-weight and $n$-distinguishability
are defined and their general properties are discussed in Sec. \ref{sec:PnThetan}.
Sec. \ref{sec:eth-ising} is devoted to analyzing numerical results
in the Ising model, while Sec. \ref{sec:conclusion} is dedicated
to further discussions and conclusion.

\section{Fock space of operators\label{sec:fock-space}}

We start by defining a mapping from a lattice many-body system to
a single-particle quantum mechanics problem on a high-dimensional
graph. For concreteness and simplicity, let us choose the many body
system to be a spin-1/2 model with Hamiltonian $H$ on a chain with
$L$ sites. For a spin-1/2 model, the Hilbert space of site $r$ is
spanned by an orthonormal basis consisting of the identity matrix
and the Pauli matrices: $\frac{1}{\sqrt{2}}\left(\mathbb{1}_{r},\sigma_{r}^{x},\sigma_{r}^{y},\sigma_{r}^{z}\right)\equiv O_{(r,i)},i=0,1,2,3$,
normalized as $\mbox{Tr}\left(O_{(r,i)}O_{(r,j)}\right)=\delta_{ij}$.
They satisfy the orthogonality condition 
\begin{equation}
\sum_{i=1}^{3}O_{(r,i)}^{ab}O_{(r,i)}^{cd}=\delta_{ad}\delta_{bc}-\frac{1}{D}\delta_{ab}\delta_{cd}\equiv W_{r}=X_{r}-\frac{\mathbb{1}_{r}}{D}\label{eq:single-site-ortho}
\end{equation}
where $D=2$ is the dimension of the Hilbert space at each site, $X_{r}^{ab,cd}=\delta_{ad}\delta_{bc}$
is the ``swap'' operator acting on site $r$, whose expectation
value computed over two copies of any state $\rho$ is related to
the second Renyi entropy of site $r$ in that state \cite{Hastings2010}:
$S_{r}=-\log\mbox{Tr}_{\bar{r}}\left(\mbox{Tr}_{r}\rho\right)^{2}=-\log\mbox{Tr}\left[(\rho\otimes\rho) X_{r}\right]$,
and $W_{r}$ is its traceless part. Here, $\bar{r}$ denotes all sites
except site $r$. The Hilbert space of the whole system is spanned
by multisite operators $O_{\{r_{\alpha},i_{\alpha}|\alpha=1\dots n\}}^{(n)}=O_{(r_{1},i_{1})}\otimes O_{(r_{2},i_{2})}\dots O_{(r_{n},i_{n})}\equiv O_{\ell}^{(n)}$,
where $n$ represents the number of sites on which $O_{\ell}^{(n)}$
is one of the Pauli matrices, and sites absent in the set $\{r_{\alpha}\}$
are assumed to host the normalized identity operator $\frac{\mathbb{1}}{\sqrt{2}}$
in $O_{\ell}^{(n)}$. Henceforth, $n$ will be referred to as the
\emph{size} of the operator $O_{\ell}^{(n)}$. The subscript $\ell$
indexes all operators with size $n$, so it runs from $1$ through
$(D^{2}-1)^{n}\left(\begin{array}{c}
L\\
n
\end{array}\right)\equiv f_{n}D^{2L}$. We will refer to operators with small (large) $n$ as simple (complicated)
operators.

\begin{figure}
\begin{centering}
\includegraphics[width=0.8\columnwidth]{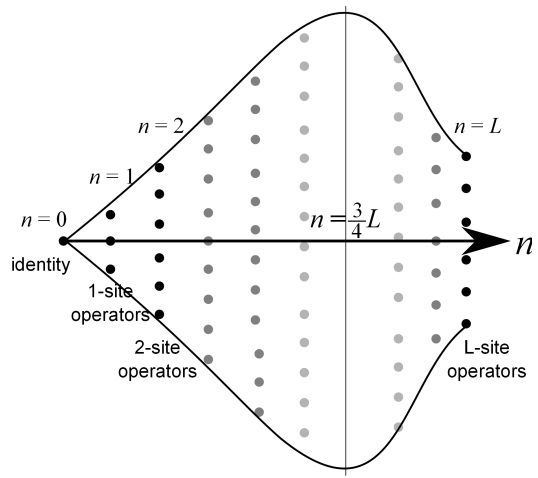} 
\par\end{centering}

\caption{Schematic of the tree network, constructed by arranging operators
according to their size. The identity operator is on the extreme left,
followed by the single-site operators, and so on. The total number
of operators is $D^{2L}$, while the fraction in the $n^{th}$ layer
is $f_{n}=\frac{(D^{2}-1)^{n}}{D^{2L}}\left(\protect\begin{array}{c}
L\protect\\
n
\protect\end{array}\right)$, which is maximum for $n=(1-1/D^{2})L=\frac{3}{4}L\equiv n^{*}$
for $D=2$. Note that $f_{n}$ grows exponentially with $n$ for small
$n$, so the apparent linear growth in the number of dots with $n$
in the figure is for ease of depiction and should not be taken literally.\label{fig:tree-schematic}}
\end{figure}

Now, we construct a graph by assigning a node to each basis 
operator $O_{\ell}^{(n)}$. The nodes are sorted by operator-size,
so the $n^{th}$ layer of the graph is comprised of nodes corresponding
to $n$-site operators, as depicted in Fig. \ref{fig:tree-schematic}.
The total number of operators, or nodes in the graph, is $D^{2L}$,
while the number of nodes in each layer is $f_{n}D^{2L}$. The outline
in Fig. \ref{fig:tree-schematic} depicts the number of nodes in each
layer, $f_{n}D^{2L}$. For $L\gg D^{2}$, $f_{n}$ is maximized at
$n=n^{*}=(1-1/D^{2})L$ which equals $3L/4$ for $D=2$. It is easy
to check that $n^{*}$ is the mean, median as well as the mode
of $f_{n}$. Physically, this means that a randomly chosen basis operator
is most likely to have a size of $n^{*}$, since each site has $D^{2}-1$
non-trivial operators and a single trivial one.

The orthogonality of the operator basis allows us to expand an arbitrary
many-body Hermitian operator $A(t)$ in this basis as $A(t)=\sum_{n,\ell}\psi_{n\ell}(A;t)O_{\ell}^{(n)}$,
with real coefficients:

\begin{equation}
\psi_{n\ell}(A;t)=\mbox{Tr}\left(A(t)O_{\ell}^{(n)}\right)\label{eq:wavefunction-def}
\end{equation}
The vector $\psi_{n\ell}(A;t)$ can be viewed as a "single-particle
wavefunction" of a particle hopping on the graph we defined. There
are two key requirements for interpreting $\psi_{n\ell}(A;t)$ as
a sensible single-particle wavefunction within first quantization.
Firstly, the total probability density of the particle must be conserved;
this is guaranteed because $\left\{ O_{\ell}^{(n)}\right\} $ form
an orthonormal basis for operators in the Hilbert space, which implies
\begin{equation}
\sum_{n,\ell}\left|\psi_{n\ell}(A;t)\right|^{2}=\mbox{Tr}\left(A^2\right)\label{eq:psi-norm}
\end{equation}
a manifestly invariant quantity under unitary time-evolution of $A$.
Secondly, it must satisfy Schrodinger's equation. Indeed, $A(t)$
follows Heisenberg time evolution: $\dot{A}(t)=i[H,A(t)]$, so the
time-evolution of $\psi_{n\ell}$ is given by

\begin{eqnarray}
i\frac{\partial}{\partial t}\psi_{n\ell}(A;t) & = & \mbox{Tr}\left(\left[A(t),H\right]O_{\ell}^{(n)}\right)\nonumber \\
 & = & \mathcal{H}_{\ell\ell'}^{nn'}\psi_{n'\ell'}(A;t)\label{eq:Schrodinger}
\end{eqnarray}
where the hopping matrix element between two nodes $O_{\ell}^{(n)}$
and $O_{\ell'}^{(n')}$ is defined as 
\begin{equation}
\mathcal{H}_{\ell\ell'}^{nn'}=\mbox{Tr}\left(\left[H,O_{\ell}^{(n)}\right]O_{\ell'}^{(n')}\right)=-\mathcal{H}_{\ell'\ell}^{n'n}=-\left(\mathcal{H}_{\ell\ell'}^{nn'}\right)^{*}\label{eq:hopping}
\end{equation}
Thus, $\psi_{n\ell}(A;t)$ is a sensible wavefunction, and its dynamics
are governed by the hopping Hamiltonian (\ref{eq:hopping}). Crucially,
if the physical Hamiltonian contains at most $k$-spin terms, $\mathcal{H}_{\ell\ell'}^{nn'}$
vanishes for $|n-n'|\ge k$ and therefore, satisfies a notion of locality
on the graph. Thus, the time-evolution of a general operator in a
many body system has been recast into the problem of a single particle
governed by a local hopping Hamiltonian on a high-dimensional graph.
The ``energies'' of the particle consists of all possible differences
$E_{i}-E_{j};i,j=1\dots D^{L}$ between pairs of energies of $H$,
so the single-particle spectrum that results from a generic $H$ with
no degeneracies is particle-hole symmetric with $D^{L}$ zero eigenvalues
and $D^{2L}-D^{L}$ non-zero ones. The above construction is reminiscent
of mappings of states in Fock space to a Cayley tree, which allows
one to view integrable systems with local conservation laws in real
\cite{Altshuler1997} and momentum \cite{Neuenhahn2012} space as
a localized particle on a suitably defined tree. Our construction,
on the other hand, describes the Fock space of operators instead of
that of states, and thus is closer to the approach adopted by Ref. \cite{Ros2015} for constructing integrals of motion to the describe the many-body localized phase.

The graph construction facilitates extracting different kinds of information
about quantum ergodicity with different choices of $A$. If we choose
$A$ to be the density matrix $\rho$ for an eigenstate that satisfies
the ETH, then $\psi_{n\ell}(A)$ is expected to coincide with its
thermal value, determined only by the energy density in that state, for
small $n$, but can be depend on other details for larger $n$. On
the other hand, picking $A$ to be the difference between two density
matrices tells us how various correlators differ in the two states,
and what kind of observables must be measured in order to distinguish
between them. Finally, if $A$ is a physical observable $O$, its
time-evolution tells us how this physical quantity evolves into a
superposition of other operators with time. In particular, suppose
we start from a small-$n$ operator and let it evolve in time. In
general, it will evolve into a superposition containing many large-$n$
operators. Equivalently, one can think of $\psi_{n\ell}(O;t)$ as
an infinite temperature correlation function, so its time evolution
tells us how small operators develop correlations with large operators
over time. In the language of the hopping particle, this corresponds
to the particle starting near the low-$n$ end of the graph and spreading
towards larger $n$ nodes. Thus, chaotic behavior of operators turns
into delocalization of the particle on the graph.

\section{$n$-weight and $n$-distinguishability\label{sec:PnThetan}}

A central quantity that we will work with in this paper is the $n$-weight
of an operator $A$, $P_{n}(A)$, defined as 
\begin{equation}
P_{n}(A)=\sum_{\ell}\left|\psi_{n\ell}(A)\right|^{2}=\sum_{\{r_{\alpha},i_{\alpha}\}}\left|\mbox{Tr}\left(AO_{\{r_{\alpha},i_{\alpha}|\alpha=1\dots n}\right)\right|^{2}\label{eq:Pn-def}
\end{equation}
where the sum over $\ell$ runs over all possible choices of $n$-site
operators. In the graph picture of Sec. \ref{sec:fock-space}, this
is the total probability density in the $n^{th}$ layer of the graph.
Physically, $P_{n}(A)$ tells us how complex the correlators one must
measure in order to reconstruct $A$ are. Computing it directly is
computationally taxing, as it entails computing $D^{2L}$ traces,
one for each operator in the Hilbert space. Fortunately, the computation
can be simplified via a generating function, as follows.

Performing the sums over $\{i_{\alpha}\}$ for fixed $\{r_{\alpha}\}$
and using Eq. (\ref{eq:single-site-ortho}) gives 
\begin{equation}
P_{n}(A)=\frac{1}{D^{L-n}}\sum_{R_{n}}\prod_{r\in R_{n}}\mbox{Tr}\left[(A\otimes A)W_{r}\right]\label{eq:PnW}
\end{equation}
where $R_{n}$ is a region of size $n$ (not necessarily connected),
and the sum $\sum_{R_{n}}$ is over all $n$-site regions. Now we
define a generating function 
\begin{eqnarray}
F(A;z)=\sum_{n=0}^{L}z^{n}P_{n}(A)
\end{eqnarray}
which can be explicitly written as 
\begin{eqnarray}
F(A;z) & = & \frac{1}{D^{L}}\mbox{Tr}\left[A\otimes A\prod_{r=1}^{L}\left(\mathbb{1}_{r}+DzW_{r}\right)\right]\label{eq:gen-func}\\
 & = & \frac{1}{D^{L}}\sum_{R}(1-z)^{L-k_{R}}(Dz)^{k_{R}}\mbox{Tr}_{R}\left(\mbox{Tr}_{\bar{R}}A\right)^{2}\nonumber 
\end{eqnarray}
Here the sum $\sum_{R}$ is over all regions $R$, composed of $k_{R}$
sites and $\bar{R}$ denotes the complement of $R$. $P_{n}(A)$ is
then determined by Fourier transforming $F(A;e^{in\theta})$: 
\begin{equation}
P_{n}(A)=\frac{1}{2\pi}\int\mathrm{d}\theta e^{-in\theta}F(A;e^{in\theta})
\end{equation}
Since $n\in[0,L]$ is linear in system size, it is actually sufficient
to calculate $F(A;e^{in\theta})$ for the $L+1$ discrete values of
$\theta=\frac{2\pi}{L+1}m,~m=0,1,..,L$. Therefore we have translated
the calculation of $P_{n}(A)$ for all $A$ to $L+1$ operator trace
computations in the doubled Hilbert space. Alternatively, we can also
use the second line of Eq. (\ref{eq:gen-func}) and calculate $P_{n}$
by performing $D^{L}$ partial trace calculations on a single copy
of the system.

When $A$ is a density matrix $\rho$, the generating function also
makes it transparent that $P_{n}(A)$ is related to the second Renyi
entropy of a region $R$, $S_{R}=-\log{\rm Tr}\left(\rho_{R}\right)^{2}=-\log\mbox{Tr}\left[\rho\otimes\rho\prod_{r\in R}X_{r}\right]$.
From (\ref{eq:PnW}), we have 
\begin{equation}
P_{n}(\rho)=\frac{1}{D^{L-n}}\sum_{R_{n}}\sum_{R_{k}\subseteq R_{n}}e^{-S_{R_{k}}}\left(-\frac{1}{D}\right)^{n-k}\label{eq:Pn-explicit1}
\end{equation}
The sum over $R_{n}$ and $R_{k}$ can be combined into a single sum
by introducing suitable combinatorial factors. Defining $e^{-\mathbb{S}_{k}}=\left\langle e^{-S_{R_{k}}}\right\rangle _{R_{k}}$,
i.e., the average of $e^{-S_{R_{k}}}$ over all $k$-site regions,
we get 
\begin{equation}
P_{n}(\rho)=\frac{1}{D^{L-n}}\left(\begin{array}{c}
L\\
n
\end{array}\right)\sum_{k=0}^{n}e^{-\mathbb{S}_{k}}\left(-\frac{1}{D}\right)^{n-k}\left(\begin{array}{c}
n\\
k
\end{array}\right)\label{eq:Pn2Sn}
\end{equation}
Thus, there is a simple relationship between $P_{n}$, the ``single
particle density on the graph'', and entanglement properties of the
many body state.

\begin{figure}
\begin{centering}
\includegraphics[width=0.48\columnwidth]{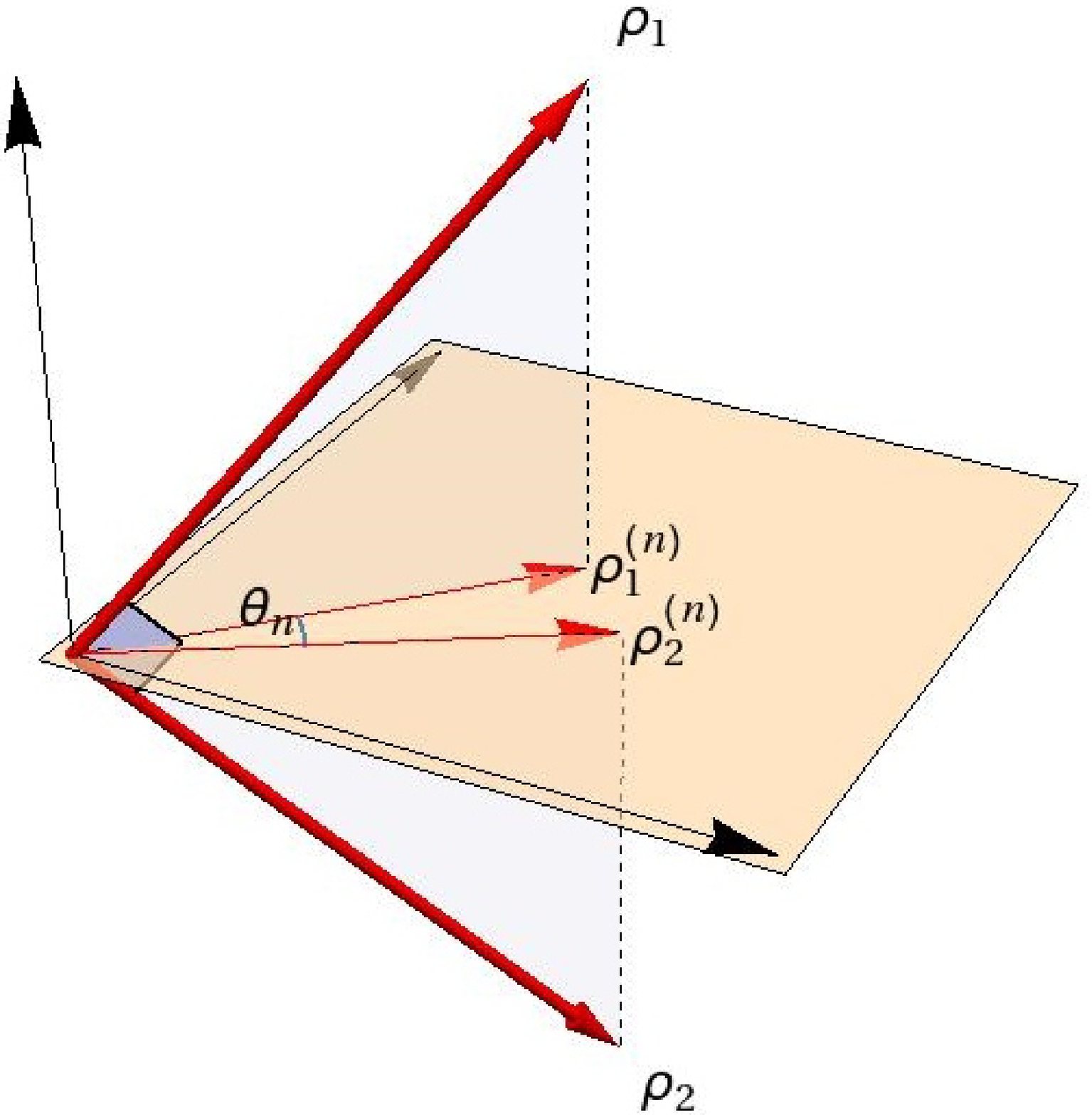}\includegraphics[width=0.48\columnwidth]{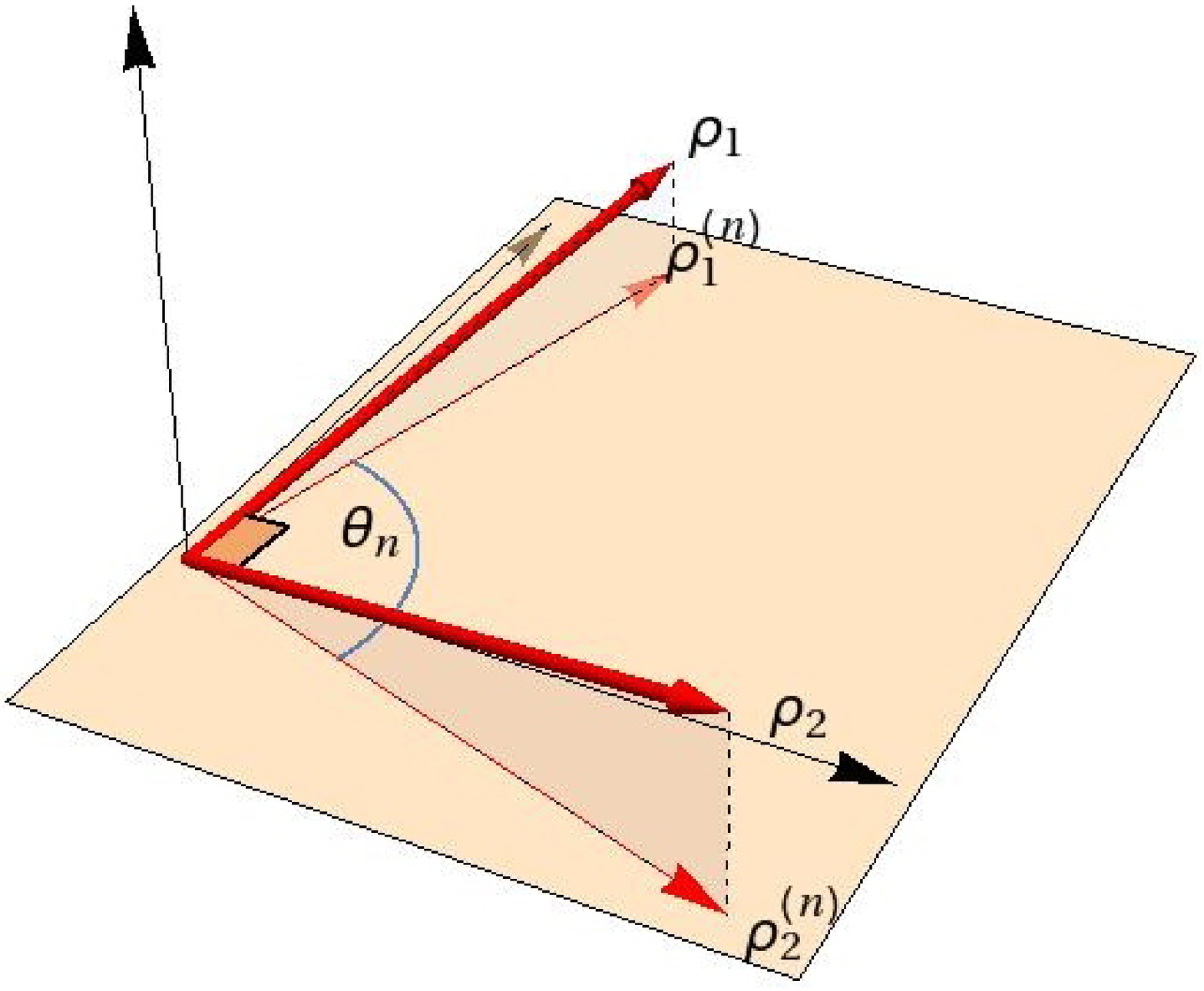} 
\par\end{centering}

\caption{Schematic illustration of the behavior of neighboring eigenstates
$\rho_{1}$ and $\rho_{2}$ on being projected onto the space of small-$n$
(left) and large-$n$ (right) operators. $\rho_{i}^{(n)}$ is shorthand
for the projections $\vec{\psi}_{n}(\rho_{i})$ defined in the text.
The three directions together depict the full Hilbert space of operators,
while the horizontal plane represents its projection onto the space of
operators of size $n$. $\rho_{1}$ and $\rho_{2}$ are mutually orthogonal
vectors in the full space. However, they appear nearly identical when
projected onto simple operators as shown on the left, but look quite
different for larger operators as shown on the right. The angle $\theta_{n}$
will be calculated numerically for the non-integrable Ising model
in Sec. \ref{sec:eth-ising}.\label{fig:thetan-sketch}}
\end{figure}

Based on the $n$-weight defined for each operator $A$, we define
a second quantity, the $n$-distinguishability between two operators
$A_{1}$ and $A_{2}$, as

\begin{equation}
\theta_{n}(A_{1},A_{2})=\cos^{-1}\frac{P_{n}(A_{1})+P_{n}(A_{2})-P_{n}(A_{1}-A_{2})}{2\sqrt{P_{n}(A_{1})P_{n}(A_{2})}}\label{eq:thetan-def}
\end{equation}
$\theta_{n}\left(A_{1},A_{2}\right)$ is simply the angle between
the two vectors $\vec{\psi}_{n}(A_{1})=\psi_{n\boldsymbol{\ell}}(A_{1})={\rm Tr}\left(A_{1}O_{\boldsymbol{\ell}}^{(n)}\right)$
and $\vec{\psi}_{n}(A_{2})$ defined similarly, i.e. the two ``single
particle wavefunctions" corresponding to $A_{1}$ and
$A_{2}$, projected to the graph sites corresponding to size-$n$
operators. The angle $\theta_{n}$ thus measures how different the
two operators are if only $n$-site operators are measured. A small
$\theta_{n}$ implies that $A_{1}$ and $A_{2}$ look similar in all
size-$n$ measurements, while a large $\theta_{n}\sim\frac{\pi}{2}$
means $A_{1}$ and $A_{2}$ can be easily distinguished by $n$-site
operators. This is sketched in Fig. \ref{fig:thetan-sketch}, where
we have chosen $A_{1,2}$ to be two density matrices $\rho_{1,2}$
in anticipation of the discussion in the next section. In Sec. \ref{sub:thetan},
we will use $\theta_{n}$ to distinguish between neighboring eigenstates
and show that indeed, they appear similar for simple observables and
different for complicated ones.

\section{eigenstate thermalization in the Ising model\label{sec:eth-ising}}

\begin{figure}
\begin{centering}
\includegraphics[width=0.49\columnwidth]{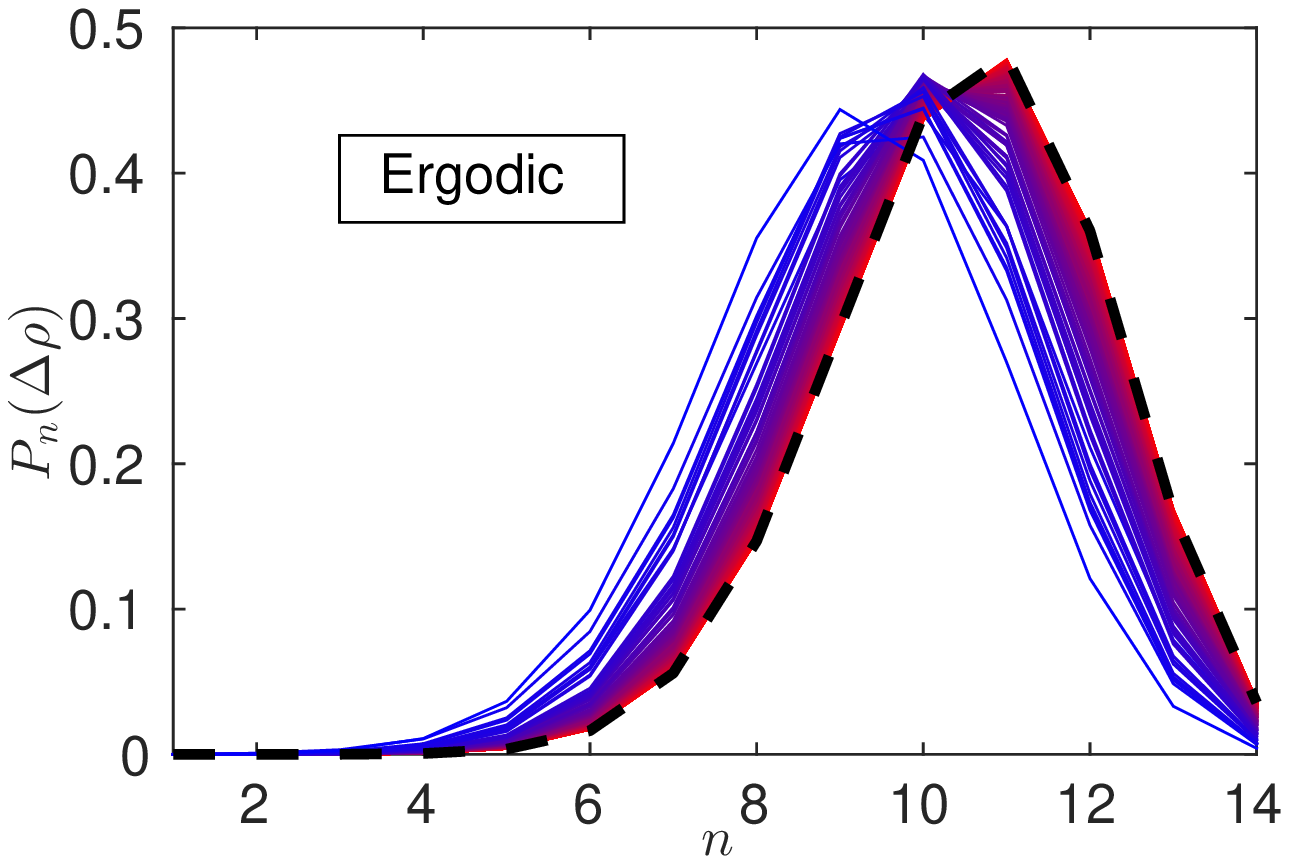}\includegraphics[width=0.49\columnwidth]{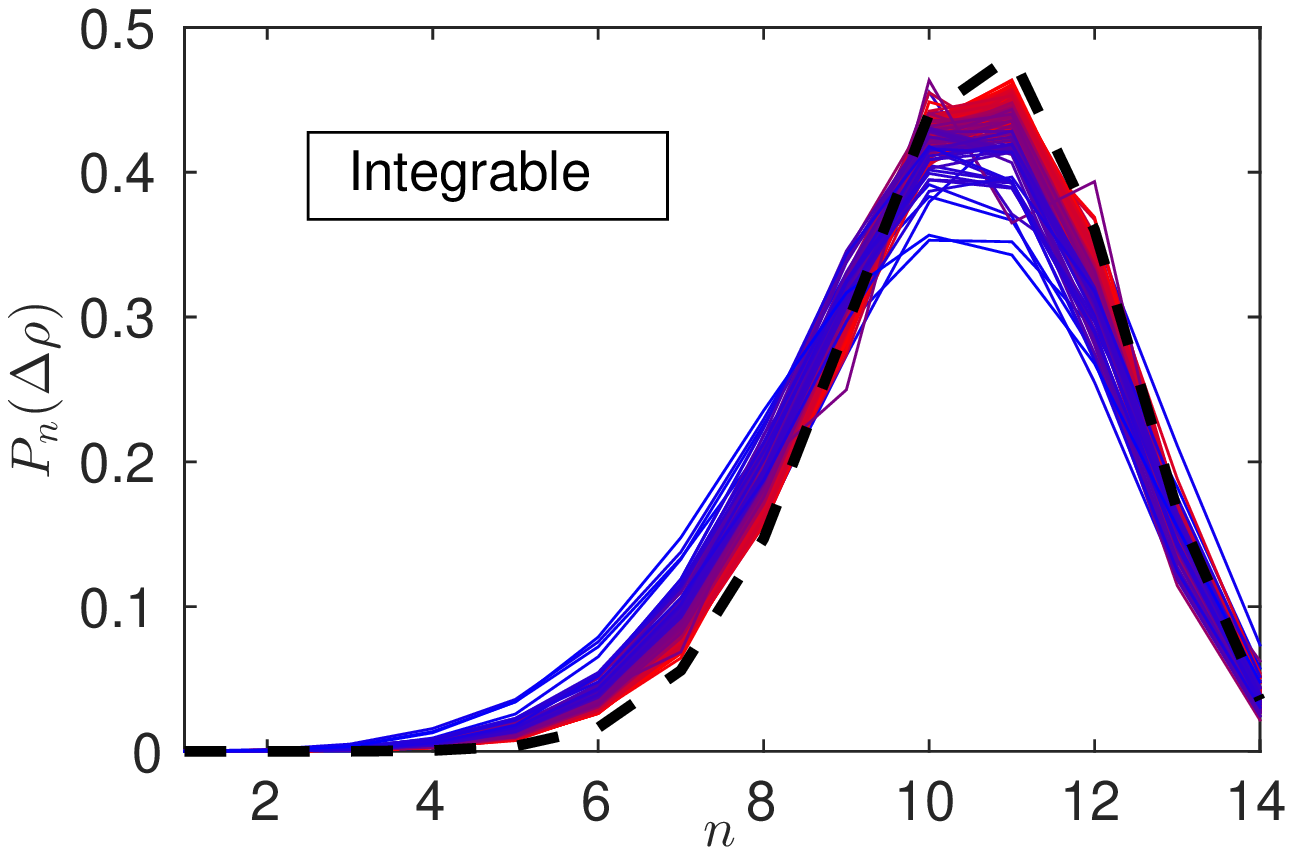} 
\par\end{centering}

\begin{centering}
\includegraphics[width=0.49\columnwidth]{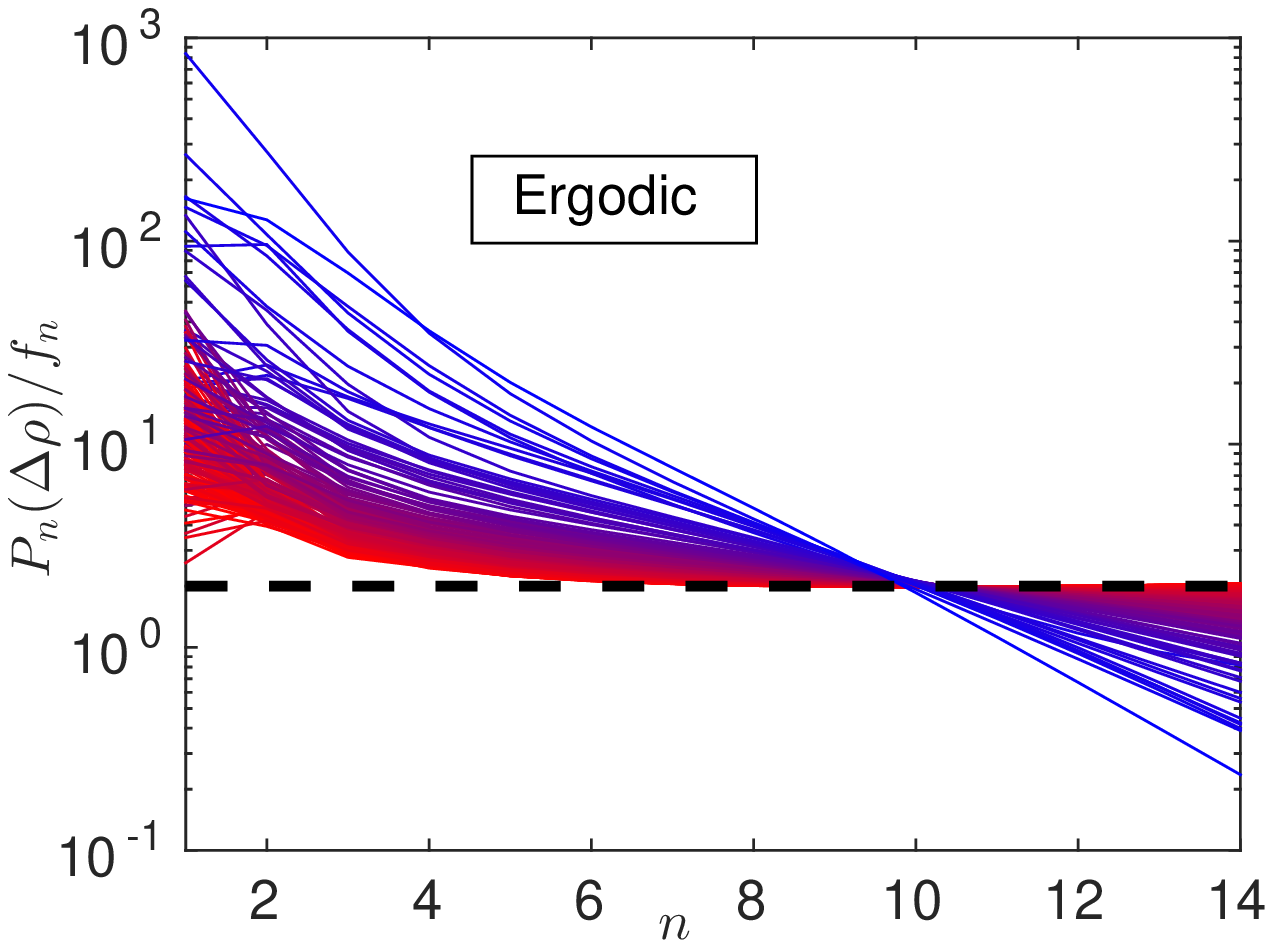}\includegraphics[width=0.49\columnwidth]{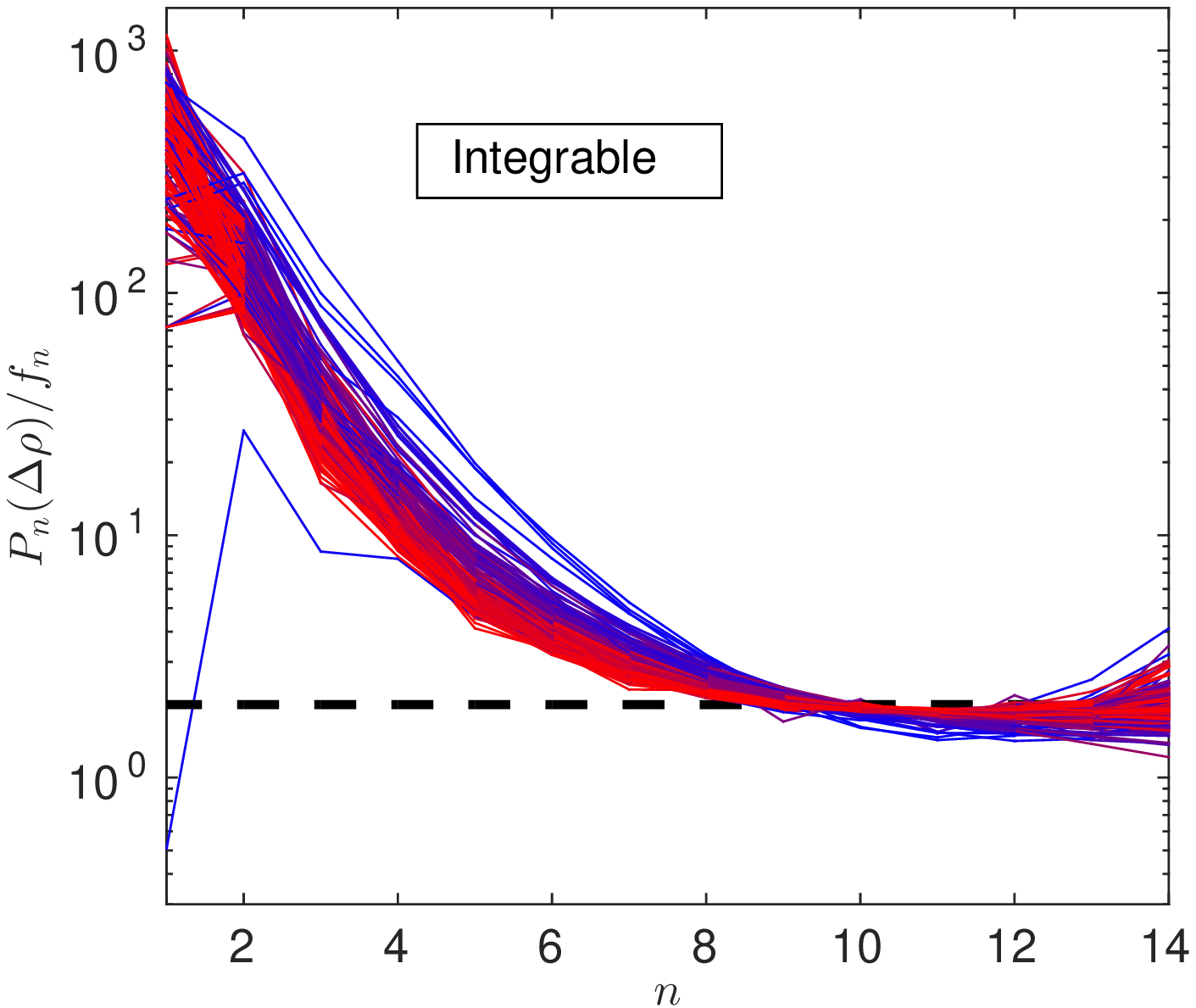} 
\par\end{centering}

\caption{$P_{n}(\Delta\rho)$ vs $n$ (above) and $P_{n}(\Delta\rho)/f_{n}$
vs $n$ (below) for $\sim200$ randomly chosen pairs of neighboring
eigenstates for $L=14$ sites for the ergodic (left) and the integrable
(right) Ising model. Here, $\Delta\rho = \rho_1-\rho_2$ is the difference between the density matrices of the two eigenstates. The color is proportional to the density of states,
with blue (red) representing states in regions of the spectrum with
low (high) density of states. The black dashed line marks $P_{n}(\Delta\rho)=2f_{n}$.\label{fig:Pn-delta-rho}}
\end{figure}

In this section we apply the new measures we define to study eigenstate
thermalization in a prototypical non-integrable spin model, namely,
the 1D Ising model with transverse and longitudinal fields, given
by 
\begin{equation}
H=\sum_{r}\left(J\sigma_{r}^{z}\sigma_{r+1}^{z}+h_{x}\sigma_{r}^{x}+h_{z}\sigma_{r}^{z}\right)+h_{z}\sigma_{1}^{z}\label{eq:Hamiltonian}
\end{equation}
$H$ is integrable if any one of $J$, $h_{x}$ and $h_{z}$ vanishes,
but is non-integrable otherwise. We choose $J=0.5$, $h_{x}=-0.74$
and $h_{z}=0.35$ as the non-integrable parameters, and $J=0.5$,
$h_{x}=0.35$ and $h_{z}=0$ as the integrable ones. Open boundary
conditions and the extra term on the first site, $h_{z}\sigma_{1}^{z}$,
ensure that translation and inversion symmetries are broken so that
there are no conserved quantities in the non-integrable case. This
is unlike several recent works which retained translational symmetry
and hence, conserved the total momentum \cite{Garrison2015,Kim2014,Rigol2009,Rigol2009a}.
The energies and eigenstates are obtained by exact diagonalization
of systems of upto $L=14$ sites.

\subsection{Comparison of eigenstate $n$-weights\label{sub:Pn}}

As stated in the introduction, the ETH says that the expectation values
of simple operators are equal in nearby eigenstates of chaotic Hamiltonians,
up to exponentially small corrections in the system size. This automatically
ensures that each eigenstate resembles a ``microcanonical ensemble'',
i.e., an equal admixture of nearby eigenstates, in the thermodynamic
limit and hence yields the ETH as stated in the introduction. Thus,
we first compare pairs of neighboring eigenstates $\rho_{1}$ and
$\rho_{2}$ by computing the total squared difference in the expectation
values of all operators of size $n$, 
\begin{equation}
P_{n}(\Delta\rho)=\sum_{\ell}\left(\left\langle O_{\ell}^{(n)}\right\rangle _{\rho_{1}}-\left\langle O_{\ell}^{(n)}\right\rangle _{\rho_{2}}\right)^{2}
\end{equation}
where $\Delta\rho=\rho_{1}-\rho_{2}$, and study its dependence on
$n$ and the energy of the pair. As shown in the upper panels of Fig.
\ref{fig:Pn-delta-rho}, this quantity has the anticipated behavior
for small $n$: it increases with $n$ and is larger when the density
of states is lower. A closer inspection, however, reveals that the
$n$-dependence seen here is deceptive, and cannot be used to declare
eigenstate thermalization. In particular, the curves approximately
follow the fraction of operators of size $n$, $f_{n}=\frac{(D^{2}-1)^{n}}{D^{2L}}\left(\begin{array}{c}
L\\
n
\end{array}\right)$ upto an overall proportionality constant. In fact, the ``infinite
temperature'' eigenstates -- states near the middle of the spectrum
where the density of states is highest -- have $P_{n}(\Delta\rho)\approx2f_{n}=\mbox{Tr}\left[(\Delta\rho)^{2}\right]f_{n}$.
Moreover, the $n$-dependence is roughly the same even for the integrable
Ising model. Thus, we conclude that the bare $n$-dependence of $P_{n}(\Delta\rho)$
is primarily determined by the number of operators of size $n$, not
by the integrability properties of the Hamiltonian.

Therefore we study the average density per site $P_{n}(\Delta\rho)/f_{n}$
-- the mean squared difference in the expectation values of size-$n$
operators between neighboring eigenstates, upto an overall proportionality
constant of $D^{2L}$. As is shown in the lower panels of Fig. \ref{fig:Pn-delta-rho},
the $n$-dependence of this quantity is clearly different for ergodic
and integrable systems. However, its $n$-dependence for the ergodic
system is the exact opposite of what one would naively expect from
ETH. Indeed, the lower panels of Fig. \ref{fig:Pn-delta-rho} show
that on average, large operators are actually worse at distinguishing
between neighboring eigenstates than small operators are, irrespective
of whether the Hamiltonian is integrable or not.

The fact that $P_{n}(\Delta\rho)/f_{n}$ decreases with $n$ for small
$n$ simply says that on average, simple operators store more information
about the state of the system compared to complicated ones. For
integrable systems as well as for ground states of ergodic systems,
this statement is easily understood because nearby eigenstates \emph{can
}be distinguished by simple operators. Fig. \ref{fig:Pn-delta-rho}
says that random simple operators can split finite energy density
states of ergodic Hamiltonians as well, but the efficiency with which
they can do so decreases with increasing density of states. For the
infinite temperature states (i.e., states at the part of the spectrum
with the largest density of states) $P_{n}/f_{n}$ is almost independent
of $n$, which means that there is no difference between simple and
complicated operators, since the state is essentially a random state
in the Hilbert space.

\begin{figure*}
\begin{centering}
\includegraphics[width=2\columnwidth]{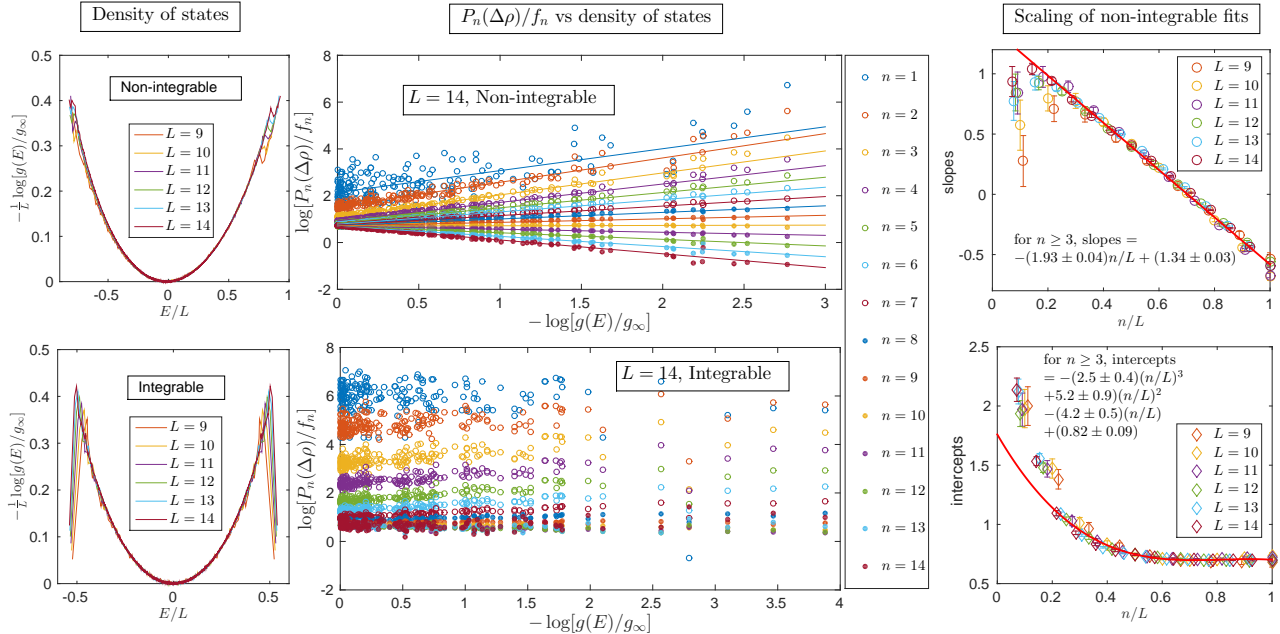} 
\par\end{centering}

\caption{Left: $\frac{1}{L}\log\left[g_{\infty}/g(E)\right]$ vs the energy
density $E/L$ for various system sizes for the ergodic (above) and
the integrable (below) Ising model. Except near the band edges, the
curves are indistinguishable, indicating that $g(E)/g_{\infty}$ grows
exponentially with $L$ with an energy density dependent exponent.
Middle: $\log\left[P_{n}(\Delta\rho)/f_{n}\right]$ vs $\log\left[g_{\infty}/g(E)\right]$
for various $n$ for $L=14$ for the ergodic (above) and the integrable
(below) Ising model, and straight line fits to the ergodic data. The
fits are good for $n\ge3$. A similar fitting procedure for other
system sizes yields the panels on the right, where we show that the
slopes (above) and the intercepts (below) of the straight lines are
simple functions of $n/L$. \label{fig:scaling}}
\end{figure*}


A peculiar feature of Fig. \ref{fig:Pn-delta-rho} is that random
operators of size $n=n^{*}\approx3L/4=(1-1/D^{2})L$ cannot distinguish
between\emph{ }any\emph{ }pair of states, irrespective of the Hamiltonian.
This can be understood heuristically as follows. Since each site has
$D^{2}-1$ non-trivial operator and a single trivial operator, a random
operator has size $n^{*}$. The concurrence of $P_{n}(\Delta\rho)/f_{n}$
curves at $n=n^{*}$ reflects the fact that measuring a random operator
does not reveal any information about the state of the system. Random
operators with $n>n^{*}$ can again distinguish between neighboring
eigenstates. Unlike simple operators, however, the efficiency with
which they can do so in ergodic systems increases with increasing
density of states. In other words, random operators with $n<n^{*}$
are better at splitting low lying excitations than at splitting finite
energy density states, whereas operators with $n>n^{*}$ are better
at the opposite.

Although $P_{n}/f_{n}$ decreases with $n$ for both ergodic and integrable
systems, its dependence on the density of states is clearly different,
as can be seen from Fig. \ref{fig:Pn-delta-rho}. Thus, we highlight
the difference by plotting the same data as a function of the density
of states $g(E)$ in Fig. \ref{fig:scaling}. Interestingly, for the
non-integrable model we find that both the $n$-dependence and the
density-of-states dependence can be captured by a clear scaling form
given by 
\begin{equation}
\frac{P_{n}(\Delta\rho)}{f_{n}}\approx\left(\frac{g(E)}{g_{\infty}}\right)^{a(n/n^{*}-1)}\mathcal{S}\left(\frac{n}{L}\right)\label{eq:Pn-scaling}
\end{equation}
where $g_{\infty}=\max g(E)$, $a$ is a positive constant and $\mathcal{S}(n/L)$
is a scaling function of $O(1)$ that depends only weakly on $n/L$;
for the data shown in this section it is the exponential of a simple
polynomial (See Fig. \ref{fig:scaling}, right bottom). Eq. (\ref{eq:Pn-scaling})
depends mainly on generic properties of the system such as the density
of states of the spectrum, the Hilbert space dimension on each site
and the system size. These properties are obviously common to both
integrable and ergodic systems; in fact, the scaling of $g(E)$ with
system size shows no distinction between them, as one can see in the left
panel of Fig. \ref{fig:scaling}. However, we see that integrable
systems do not have simple scaling form as (\ref{eq:Pn-scaling})
for $P_{n}$. We hypothesize that the scaling behavior in Eq. (\ref{eq:Pn-scaling})
is a generic property of non-integrable systems, and hope that this
conjecture can be tested in other systems.

\subsection{$n$-distinguishability of neighboring eigenstates\label{sub:thetan}}

\begin{figure}
\begin{centering}
\includegraphics[width=1\columnwidth]{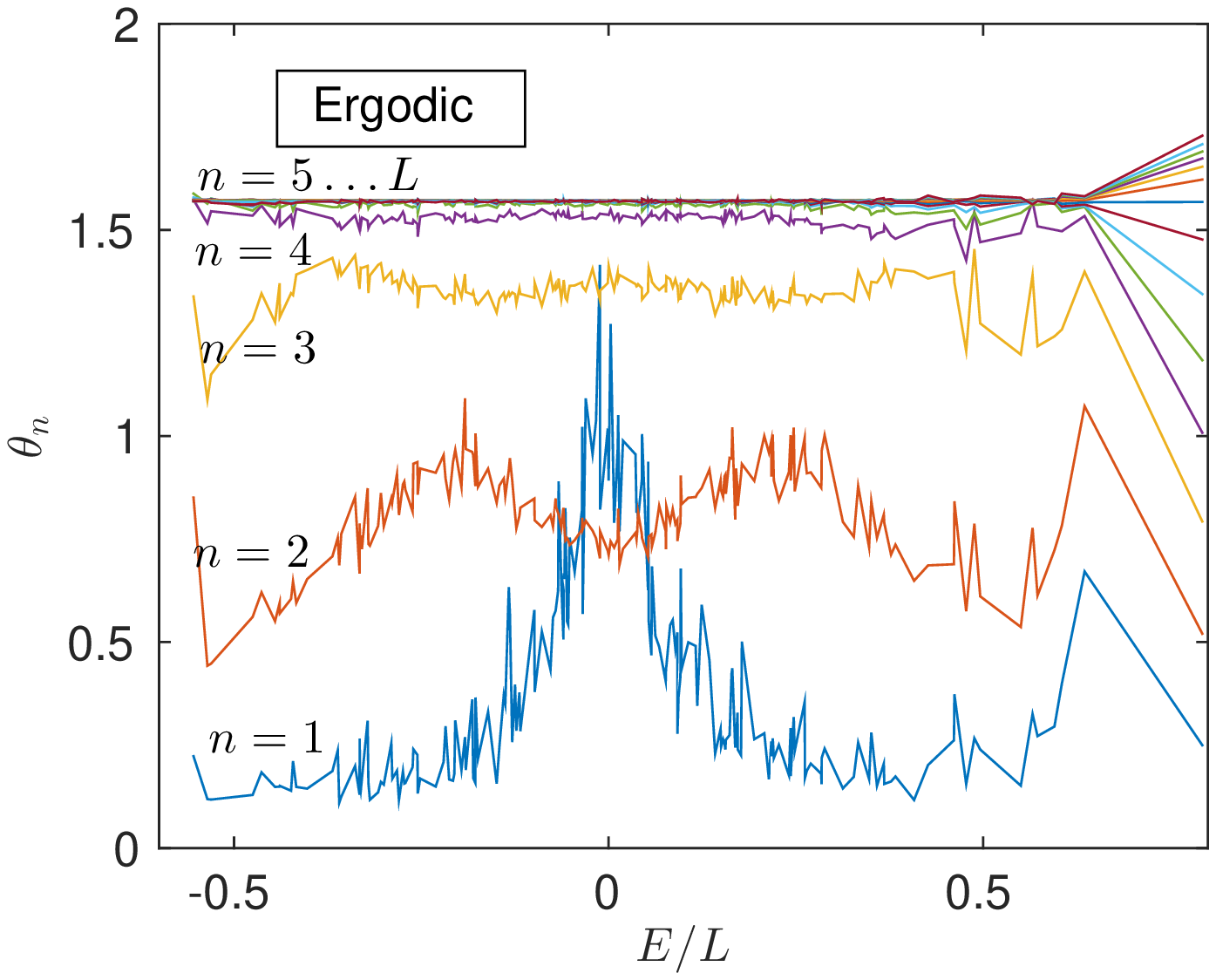} 
\par\end{centering}

\begin{centering}
\includegraphics[width=1\columnwidth]{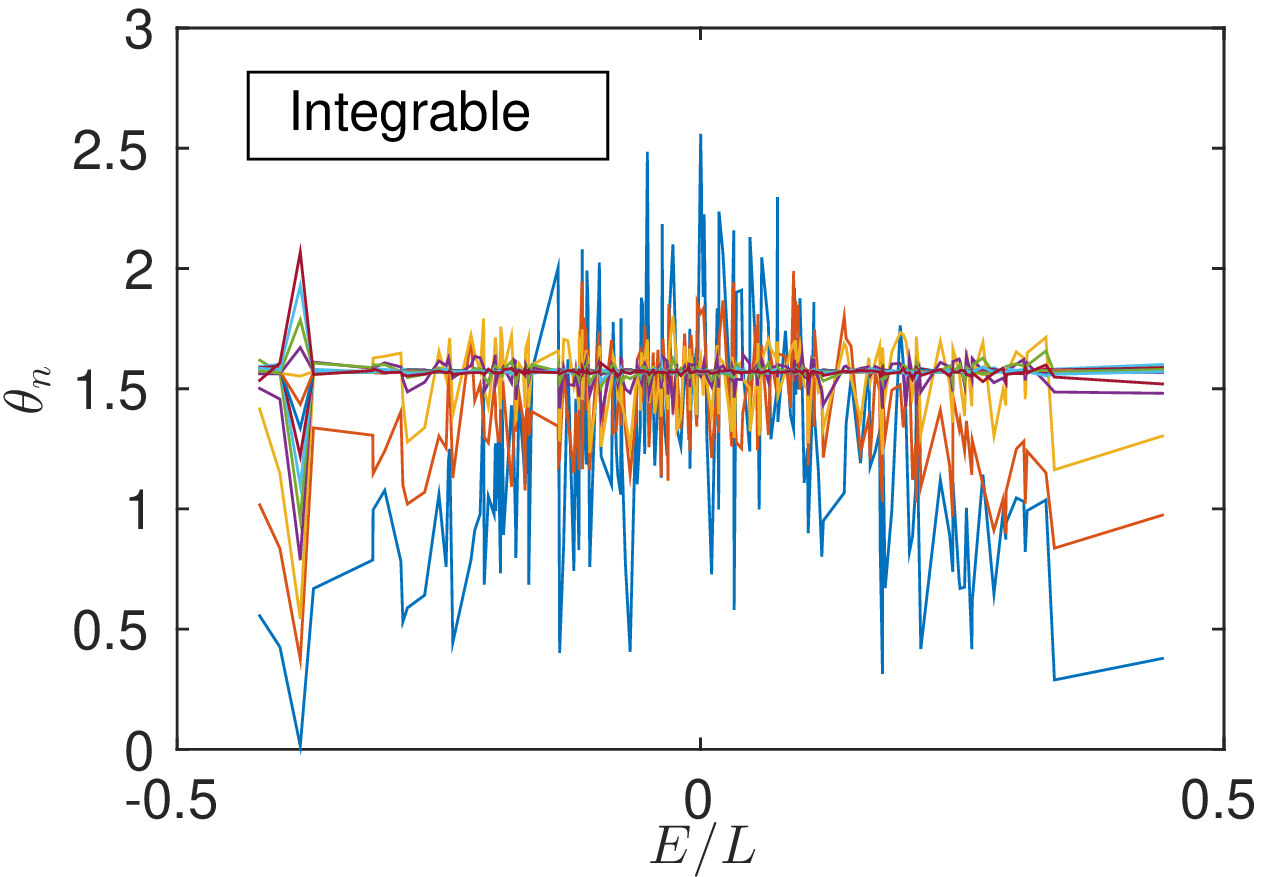} 
\par\end{centering}

\caption{The angle between the projections of two neighboring eigenstates onto
the space of operators of size $n$, $\theta_{n}$ as a function of
energy density for various $n$ for $L=14$ for the ergodic (above)
and the integrable (below) Ising model. For small $n$, $\theta_{n}$
is small for most states in the ergodic model, but is large in the
integrable one. For large $n$, it is very close to $\pi/2$ in both
figures, thus proving that neighboring eigenstates are nearly orthogonal
when projected onto the space of $n$-site operators with large $n$.
Large $\theta_{n}$ at small $n$ and large $g(E)$ ($g(E)$ is large
near the middle of the spectrum and small near the edges; see Fig.
\ref{fig:scaling}) is due to the vectors $\rho_{1}^{(n)}$ and $\rho_{2}^{(n)}$
becoming very small in magnitude; then the angle between them is meaningless.
The sharp change in $\theta_{n}$ at the lowest and\textbf{ }the highest
energies in the ergodic case originates from the fact that these states
do not satisfy the ETH at the current system size.\label{fig:thetan}}
\end{figure}

How do we reconcile the decrease of $P_{n}(\Delta\rho)/f_{n}$ with
$n$ in Fig. \ref{fig:Pn-delta-rho} with the anticipation from ETH
that expectation values of large operators, in some sense, deviate
more than those of small operators? In order to resolve this counterintuitive
behavior, we compute the $n$-distinguishability $\theta_{n}(\rho_{1},\rho_{2})$
for the same pair of neighboring eigenstates $\rho_{1}$ and $\rho_{2}$
and present the result in Fig. \ref{fig:thetan}. As we discussed
earlier, $\theta_{n}$ is the angle between two vectors in the size-$n$
Hilbert space $\vec{\psi}_n(\rho_1)$ and $\vec{\psi}_n(\rho_2)$,
and the vectors are lists of average values of all size-$n$ operators
in the two states $\rho_{1},\rho_{2}$ respectively.

As is shown in Fig. \ref{fig:thetan}, $\theta_{n}$ increases monotonically
with $n$ for most pairs of states in the ergodic system and quickly
saturates at the maximal value $\pi/2$ before $n$ reaches $L/2$.
By combining this observation with the behavior of $P_{n}/f_{n}$
observed earlier, one can understand better what happens with increasing
$n$. $P_{n}(\rho_{1}-\rho_{2})/f_{n}=|\vec{\psi}_n(\rho_{1})/\sqrt{f_{n}}-\vec{\psi}_n(\rho_{2})/\sqrt{f_{n}}|^2$
is the 2-norm squared of the difference between the two vectors,
i.e., the Euclidean distance between them. Although this distance
decreases with increasing $n$, the decrease is mainly due to the
shrinking of the norm of each vector, and the angle between them is
actually increasing. For large $n$ the two vectors are both very
short (which means a typical size-$n$ operator has a small average
value), but they are almost always exactly perpendicular. In contrast,
at small $n$ the angle is small, meaning the simple-operator average
values in the two neighboring eigenstates are well-correlated. In
other words, as $n$ increases, although a randomly chosen size-$n$
operator does a worse job distinguishing the two states, there exists
a particular choice of operator which can distinguish the two states
better. When the two vectors become perpendicular at large $n$, the
two states can be distinguished completely if we simply use the $n$-size
operator $\sum_{\ell}O_{\ell}^{(n)}\psi_{n\ell}(\Delta\rho)$, i.e.,
the projection of $\Delta\rho$ to the $n$-size subspace.

At the maximum of the density of states, at $E/L=0$, we see a peak
of $\theta_{n}$ for small $n$. The ETH is expected to work best
in this regime, but the behavior of $\theta_{n}$ indicates that it
is violated dramatically. However, this feature can be safely ignored
because at infinite temperature, the simple operators' average values
are almost vanishing so the angle between them is inconsequential.

\section{Conclusion and discussions\label{sec:conclusion}}

In summary, we proposed two related new measures of thermalization,
which help uncover a finer structure of the process of thermalization
than was previously known. By mapping the Heisenberg evolution of
operators in a quantum many-body system onto the Schrodinger evolution
of a single particle on a high dimensional graph, the complexity of
an operator can be measured by its probability distribution in the
space of all operators. We defined the $n$-weight of an operator,
which measures its weight in size-$n$ operators. We studied the behavior
of the $n$-weight for the difference between the density operator
of two neighboring eigenstates in the energy spectrum, denoted as
$P_{n}(\Delta\rho)$, which tells us how different these two states
are if we only measure size-$n$ operators. We found that $P_{n}(\Delta\rho)$
in the non-integrable Ising model follows a simple scaling behavior
as a function of the size $n$ and the density of states. In contrast
to naive expectation, we saw that large operators on average did a
\textit{worse} job at differentiating between two neighboring eigenstates
in a non-integrable system. In particular, there is a critical operator
size, which is $3/4$ of the system size for spin-$1/2$ models, at
which operators on average cannot distinguish between any states because
they themselves are ``random operators''. This counterintuitive
conclusion was explained when we investigated the
other measure of thermalization, the $n$-distinguishability, defined
as an angle between two Hilbert space vectors, each corresponding
to the projection of an operator onto the subspace of size-$n$ operators.
We saw that the angle was small for low $n$ and saturated to $\pi/2$
for large $n$, which suggests that two operators (which are orthogonal
in the whole space) are indeed almost orthogonal in the subspace of
size-$n$ with large $n$, while they are almost parallel for small
$n$. In other words, this means that although a generic size-$n$
operator does a poor job distinguishing two nearby eigenstates for
large $n$, there exists a particular choice of size-$n$ operators
which can distinguish them almost perfectly. For small size operators,
a random operator has an average value that's more different in these
two states than a random large size operator, simply because the expectation
values increase with decreasing $n$. However no small operator can
completely distinguish the two density operators since the two corresponding
vectors are almost parallel. In short, random operators are better
distinguishers for smaller $n$, but the best distinguishers are large
operators.

The ``particle-on-the-graph'' picture also allows us to understand,
heuristically, why the ETH holds in the first place. It follows from
the fact that for vectors living in a high-dimensional space
(which in this case are the states of the particle in its Hilbert
space), measuring just a few components of two vectors is insufficient
to deduce whether they are orthogonal or not, especially if the components
along some of the measured directions are almost equal. The role of
the Hamiltonian is twofold: (i) being ergodic, it ensures that the
entire space of operators is connected, as opposed to integrable systems
where it effectively decouples into subspaces with fixed values of
the conserved quantities (ii) being a sum of simple operators, it
approximately restricts the expectation values other simple operators
can take in nearby eigenstates, i.e., eigenstates with the same energy
density in the thermodynamic limit.

The intuition about the role of Hamiltonian can be further clarified
as follows. A density operator $\rho=|\phi\rangle\langle\phi|$ of
an eigenstate can be determined by the following equations: 
\begin{eqnarray}
\left[\rho,H\right] & = & 0\nonumber \\
\mbox{Tr}(\rho H^{m}) & = & E^{m},~m=1\dots D^{L}-1
\end{eqnarray}
All density operators for eigenstates can be written as a linear superposition
of the identity operator and $H$, $H^{2}\dots H^{D^{L}-1}$. If one
has access to all the coefficients of this expansion, then one can
tell that two different eigenstates are mutually orthogonal. However,
the coefficients of the low powers of $H$ are slowly varying functions
of $E$, so are almost identical for eigenstates with nearby energies.
Therefore, if we restrict the measurements to operators captured by
small powers of $H$, then it is clear why the nearby eigenstates
appear almost identical. The more nontrivial statement that requires
non-integrability of the Hamiltonian is the fact that the expectation
values of other simple operators also depend on the energy of an eigenstate
smoothly, probably because the high powers of $H$ have a negligible
contribution to simple operator average values. More rigorous analytic
and numerical work is required to make this discussion more complete,
which we will leave for the future.

Phase space based reasons for why isolated quantum systems fulfil
the basic tenets of statistical mechanics have been suggested in the
past to argue that most systems, at most times and for most observables,
behave as if they belonged to a thermal ensemble \cite{Popescu2005,Popescu2006,Linden2009}.
The proofs there were based on Levy's lemma which is the analog of
the law of large numbers for vectors in a high-dimensional space.

All the results in this work concern expectation values of operators
in eigenstates. The off-diagonal matrix elements of an operator between
different eigenstates $\langle n|O|m\rangle$ are responsible for
time-evolution properties of the system. Whether thermalization occurs
in time evolution of a state that is not an eigenstate, and how fast
it occurs, depends on two properties, the off-diagonal matrix elements
of observables, and the energy difference between eigenstates. As
long as the energy differences are sufficiently incommensurate, time
evolution causes decoherence and thus leads to equilibration \cite{Popescu2005,Popescu2006,Linden2009,Yukalov2011,Yukalov2012,Yukalov2011a}.
However, different operators may thermalize at different rates, which
is an interesting problem that can be studied using the new measures
we defined.

We would like to acknowledge insightful discussions with David A.
Huse, Alexei Kitaev, Tarun Grover, Mark Srednicki and Mathew Fisher,
and thank Andre Broido for useful comments on the draft. This work
is supported by the David and Lucile Packard foundation (PH) and the
National Science Foundation through the grant No. DMR-1151786 (XLQ).

 \bibliographystyle{apsrev}
\bibliography{therm_ref}

\begin{thebibliography}{33}
\expandafter\ifx\csname natexlab\endcsname\relax\def\natexlab#1{#1}\fi
\expandafter\ifx\csname bibnamefont\endcsname\relax
  \def\bibnamefont#1{#1}\fi
\expandafter\ifx\csname bibfnamefont\endcsname\relax
  \def\bibfnamefont#1{#1}\fi
\expandafter\ifx\csname citenamefont\endcsname\relax
  \def\citenamefont#1{#1}\fi
\expandafter\ifx\csname url\endcsname\relax
  \def\url#1{\texttt{#1}}\fi
\expandafter\ifx\csname urlprefix\endcsname\relax\def\urlprefix{URL }\fi
\providecommand{\bibinfo}[2]{#2}
\providecommand{\eprint}[2][]{\url{#2}}

\bibitem[{\citenamefont{Berry}(1977)}]{Berry1977}
\bibinfo{author}{\bibfnamefont{M.~V.} \bibnamefont{Berry}},
  \bibinfo{journal}{Journal of Physics A: Mathematical and General}
  \textbf{\bibinfo{volume}{10}}, \bibinfo{pages}{2083} (\bibinfo{year}{1977}),
  \urlprefix\url{http://stacks.iop.org/0305-4470/10/i=12/a=016}.

\bibitem[{\citenamefont{Deutsch}(1991)}]{Deutsch1991}
\bibinfo{author}{\bibfnamefont{J.~M.} \bibnamefont{Deutsch}},
  \bibinfo{journal}{Phys. Rev. A} \textbf{\bibinfo{volume}{43}},
  \bibinfo{pages}{2046} (\bibinfo{year}{1991}),
  \urlprefix\url{http://link.aps.org/doi/10.1103/PhysRevA.43.2046}.

\bibitem[{\citenamefont{Srednicki}(1994)}]{Srednicki1994}
\bibinfo{author}{\bibfnamefont{M.}~\bibnamefont{Srednicki}},
  \bibinfo{journal}{Phys. Rev. E} \textbf{\bibinfo{volume}{50}},
  \bibinfo{pages}{888} (\bibinfo{year}{1994}),
  \urlprefix\url{http://link.aps.org/doi/10.1103/PhysRevE.50.888}.

\bibitem[{\citenamefont{{Garrison} and {Grover}}(2015)}]{Garrison2015}
\bibinfo{author}{\bibfnamefont{J.~R.} \bibnamefont{{Garrison}}}
  \bibnamefont{and} \bibinfo{author}{\bibfnamefont{T.}~\bibnamefont{{Grover}}},
  \bibinfo{journal}{ArXiv e-prints}  (\bibinfo{year}{2015}),
  \eprint{1503.00729}.

\bibitem[{\citenamefont{Rigol et~al.}(2008)\citenamefont{Rigol, Dunjko, and
  Olshanii}}]{Rigol2008}
\bibinfo{author}{\bibfnamefont{M.}~\bibnamefont{Rigol}},
  \bibinfo{author}{\bibfnamefont{V.}~\bibnamefont{Dunjko}}, \bibnamefont{and}
  \bibinfo{author}{\bibfnamefont{M.}~\bibnamefont{Olshanii}},
  \bibinfo{journal}{Nature} \textbf{\bibinfo{volume}{452}},
  \bibinfo{pages}{854} (\bibinfo{year}{2008}), ISSN \bibinfo{issn}{1476-4687},
  \urlprefix\url{http://dx.doi.org/10.1038/nature06838}.

\bibitem[{\citenamefont{Rigol}(2009{\natexlab{a}})}]{Rigol2009a}
\bibinfo{author}{\bibfnamefont{M.}~\bibnamefont{Rigol}},
  \bibinfo{journal}{Phys. Rev. A} \textbf{\bibinfo{volume}{80}},
  \bibinfo{pages}{053607} (\bibinfo{year}{2009}{\natexlab{a}}),
  \urlprefix\url{http://link.aps.org/doi/10.1103/PhysRevA.80.053607}.

\bibitem[{\citenamefont{Rigol and Srednicki}(2012)}]{Rigol2012}
\bibinfo{author}{\bibfnamefont{M.}~\bibnamefont{Rigol}} \bibnamefont{and}
  \bibinfo{author}{\bibfnamefont{M.}~\bibnamefont{Srednicki}},
  \bibinfo{journal}{Physical Review Letters} \textbf{\bibinfo{volume}{108}},
  \bibinfo{pages}{110601} (\bibinfo{year}{2012}), ISSN
  \bibinfo{issn}{0031-9007},
  \urlprefix\url{http://link.aps.org/doi/10.1103/PhysRevLett.108.110601}.

\bibitem[{\citenamefont{Rigol}(2014)}]{Rigol2014}
\bibinfo{author}{\bibfnamefont{M.}~\bibnamefont{Rigol}},
  \bibinfo{journal}{Phys. Rev. Lett.} \textbf{\bibinfo{volume}{112}},
  \bibinfo{pages}{170601} (\bibinfo{year}{2014}),
  \urlprefix\url{http://link.aps.org/doi/10.1103/PhysRevLett.112.170601}.

\bibitem[{\citenamefont{Khlebnikov and Kruczenski}(2014)}]{Khlebnikov2014}
\bibinfo{author}{\bibfnamefont{S.}~\bibnamefont{Khlebnikov}} \bibnamefont{and}
  \bibinfo{author}{\bibfnamefont{M.}~\bibnamefont{Kruczenski}},
  \bibinfo{journal}{Physical Review E} \textbf{\bibinfo{volume}{90}},
  \bibinfo{pages}{050101} (\bibinfo{year}{2014}), ISSN
  \bibinfo{issn}{1539-3755},
  \urlprefix\url{http://link.aps.org/doi/10.1103/PhysRevE.90.050101}.

\bibitem[{\citenamefont{Sorg et~al.}(2014)\citenamefont{Sorg, Vidmar, Pollet,
  and Heidrich-Meisner}}]{Sorg2014}
\bibinfo{author}{\bibfnamefont{S.}~\bibnamefont{Sorg}},
  \bibinfo{author}{\bibfnamefont{L.}~\bibnamefont{Vidmar}},
  \bibinfo{author}{\bibfnamefont{L.}~\bibnamefont{Pollet}}, \bibnamefont{and}
  \bibinfo{author}{\bibfnamefont{F.}~\bibnamefont{Heidrich-Meisner}},
  \bibinfo{journal}{Phys. Rev. A} \textbf{\bibinfo{volume}{90}},
  \bibinfo{pages}{033606} (\bibinfo{year}{2014}),
  \urlprefix\url{http://link.aps.org/doi/10.1103/PhysRevA.90.033606}.

\bibitem[{\citenamefont{Marcuzzi et~al.}(2013)\citenamefont{Marcuzzi, Marino,
  Gambassi, and Silva}}]{Marcuzzi2013}
\bibinfo{author}{\bibfnamefont{M.}~\bibnamefont{Marcuzzi}},
  \bibinfo{author}{\bibfnamefont{J.}~\bibnamefont{Marino}},
  \bibinfo{author}{\bibfnamefont{A.}~\bibnamefont{Gambassi}}, \bibnamefont{and}
  \bibinfo{author}{\bibfnamefont{A.}~\bibnamefont{Silva}},
  \bibinfo{journal}{Phys. Rev. Lett.} \textbf{\bibinfo{volume}{111}},
  \bibinfo{pages}{197203} (\bibinfo{year}{2013}),
  \urlprefix\url{http://link.aps.org/doi/10.1103/PhysRevLett.111.197203}.

\bibitem[{\citenamefont{Kim et~al.}(2014)\citenamefont{Kim, Ikeda, and
  Huse}}]{Kim2014}
\bibinfo{author}{\bibfnamefont{H.}~\bibnamefont{Kim}},
  \bibinfo{author}{\bibfnamefont{T.~N.} \bibnamefont{Ikeda}}, \bibnamefont{and}
  \bibinfo{author}{\bibfnamefont{D.~A.} \bibnamefont{Huse}},
  \bibinfo{journal}{Phys. Rev. E} \textbf{\bibinfo{volume}{90}},
  \bibinfo{pages}{052105} (\bibinfo{year}{2014}),
  \urlprefix\url{http://link.aps.org/doi/10.1103/PhysRevE.90.052105}.

\bibitem[{\citenamefont{Khemani et~al.}(2014)\citenamefont{Khemani, Chandran,
  Kim, and Sondhi}}]{Khemani2014}
\bibinfo{author}{\bibfnamefont{V.}~\bibnamefont{Khemani}},
  \bibinfo{author}{\bibfnamefont{A.}~\bibnamefont{Chandran}},
  \bibinfo{author}{\bibfnamefont{H.}~\bibnamefont{Kim}}, \bibnamefont{and}
  \bibinfo{author}{\bibfnamefont{S.~L.} \bibnamefont{Sondhi}},
  \bibinfo{journal}{Phys. Rev. E} \textbf{\bibinfo{volume}{90}},
  \bibinfo{pages}{052133} (\bibinfo{year}{2014}),
  \urlprefix\url{http://link.aps.org/doi/10.1103/PhysRevE.90.052133}.

\bibitem[{\citenamefont{Alba}(2015)}]{Alba2015}
\bibinfo{author}{\bibfnamefont{V.}~\bibnamefont{Alba}}, \bibinfo{journal}{Phys.
  Rev. B} \textbf{\bibinfo{volume}{91}}, \bibinfo{pages}{155123}
  (\bibinfo{year}{2015}),
  \urlprefix\url{http://link.aps.org/doi/10.1103/PhysRevB.91.155123}.

\bibitem[{\citenamefont{Beugeling et~al.}(2015)\citenamefont{Beugeling,
  Moessner, and Haque}}]{Beugeling2015}
\bibinfo{author}{\bibfnamefont{W.}~\bibnamefont{Beugeling}},
  \bibinfo{author}{\bibfnamefont{R.}~\bibnamefont{Moessner}}, \bibnamefont{and}
  \bibinfo{author}{\bibfnamefont{M.}~\bibnamefont{Haque}},
  \bibinfo{journal}{Phys. Rev. E} \textbf{\bibinfo{volume}{91}},
  \bibinfo{pages}{012144} (\bibinfo{year}{2015}),
  \urlprefix\url{http://link.aps.org/doi/10.1103/PhysRevE.91.012144}.

\bibitem[{\citenamefont{Beugeling et~al.}(2014)\citenamefont{Beugeling,
  Moessner, and Haque}}]{Beugeling2014}
\bibinfo{author}{\bibfnamefont{W.}~\bibnamefont{Beugeling}},
  \bibinfo{author}{\bibfnamefont{R.}~\bibnamefont{Moessner}}, \bibnamefont{and}
  \bibinfo{author}{\bibfnamefont{M.}~\bibnamefont{Haque}},
  \bibinfo{journal}{Phys. Rev. E} \textbf{\bibinfo{volume}{89}},
  \bibinfo{pages}{042112} (\bibinfo{year}{2014}),
  \urlprefix\url{http://link.aps.org/doi/10.1103/PhysRevE.89.042112}.

\bibitem[{\citenamefont{Ikeda et~al.}(2011)\citenamefont{Ikeda, Watanabe, and
  Ueda}}]{Ikeda2011}
\bibinfo{author}{\bibfnamefont{T.~N.} \bibnamefont{Ikeda}},
  \bibinfo{author}{\bibfnamefont{Y.}~\bibnamefont{Watanabe}}, \bibnamefont{and}
  \bibinfo{author}{\bibfnamefont{M.}~\bibnamefont{Ueda}},
  \bibinfo{journal}{Phys. Rev. E} \textbf{\bibinfo{volume}{84}},
  \bibinfo{pages}{021130} (\bibinfo{year}{2011}),
  \urlprefix\url{http://link.aps.org/doi/10.1103/PhysRevE.84.021130}.

\bibitem[{\citenamefont{Lubkin}(1978)}]{Lubkin1978}
\bibinfo{author}{\bibfnamefont{E.}~\bibnamefont{Lubkin}},
  \bibinfo{journal}{Journal of Mathematical Physics}
  \textbf{\bibinfo{volume}{19}}, \bibinfo{pages}{1028} (\bibinfo{year}{1978}),
  ISSN \bibinfo{issn}{00222488},
  \urlprefix\url{http://scitation.aip.org/content/aip/journal/jmp/19/5/10.1063/1.523763}.

\bibitem[{\citenamefont{Page}(1993)}]{Page1993}
\bibinfo{author}{\bibfnamefont{D.~N.} \bibnamefont{Page}},
  \bibinfo{journal}{Phys. Rev. Lett.} \textbf{\bibinfo{volume}{71}},
  \bibinfo{pages}{1291} (\bibinfo{year}{1993}),
  \urlprefix\url{http://link.aps.org/doi/10.1103/PhysRevLett.71.1291}.

\bibitem[{\citenamefont{Foong and Kanno}(1994)}]{Foong1994}
\bibinfo{author}{\bibfnamefont{S.~K.} \bibnamefont{Foong}} \bibnamefont{and}
  \bibinfo{author}{\bibfnamefont{S.}~\bibnamefont{Kanno}},
  \bibinfo{journal}{Phys. Rev. Lett.} \textbf{\bibinfo{volume}{72}},
  \bibinfo{pages}{1148} (\bibinfo{year}{1994}),
  \urlprefix\url{http://link.aps.org/doi/10.1103/PhysRevLett.72.1148}.

\bibitem[{\citenamefont{S\'anchez-Ruiz}(1995)}]{Sanchez-Ruiz1995}
\bibinfo{author}{\bibfnamefont{J.}~\bibnamefont{S\'anchez-Ruiz}},
  \bibinfo{journal}{Phys. Rev. E} \textbf{\bibinfo{volume}{52}},
  \bibinfo{pages}{5653} (\bibinfo{year}{1995}),
  \urlprefix\url{http://link.aps.org/doi/10.1103/PhysRevE.52.5653}.

\bibitem[{\citenamefont{Sen}(1996)}]{Sen1996}
\bibinfo{author}{\bibfnamefont{S.}~\bibnamefont{Sen}}, \bibinfo{journal}{Phys.
  Rev. Lett.} \textbf{\bibinfo{volume}{77}}, \bibinfo{pages}{1}
  (\bibinfo{year}{1996}),
  \urlprefix\url{http://link.aps.org/doi/10.1103/PhysRevLett.77.1}.

\bibitem[{\citenamefont{Hastings et~al.}(2010)\citenamefont{Hastings,
  Gonz\'alez, Kallin, and Melko}}]{Hastings2010}
\bibinfo{author}{\bibfnamefont{M.~B.} \bibnamefont{Hastings}},
  \bibinfo{author}{\bibfnamefont{I.}~\bibnamefont{Gonz\'alez}},
  \bibinfo{author}{\bibfnamefont{A.~B.} \bibnamefont{Kallin}},
  \bibnamefont{and} \bibinfo{author}{\bibfnamefont{R.~G.} \bibnamefont{Melko}},
  \bibinfo{journal}{Phys. Rev. Lett.} \textbf{\bibinfo{volume}{104}},
  \bibinfo{pages}{157201} (\bibinfo{year}{2010}),
  \urlprefix\url{http://link.aps.org/doi/10.1103/PhysRevLett.104.157201}.

\bibitem[{\citenamefont{Altshuler et~al.}(1997)\citenamefont{Altshuler, Gefen,
  Kamenev, and Levitov}}]{Altshuler1997}
\bibinfo{author}{\bibfnamefont{B.~L.} \bibnamefont{Altshuler}},
  \bibinfo{author}{\bibfnamefont{Y.}~\bibnamefont{Gefen}},
  \bibinfo{author}{\bibfnamefont{A.}~\bibnamefont{Kamenev}}, \bibnamefont{and}
  \bibinfo{author}{\bibfnamefont{L.~S.} \bibnamefont{Levitov}},
  \bibinfo{journal}{Phys. Rev. Lett.} \textbf{\bibinfo{volume}{78}},
  \bibinfo{pages}{2803} (\bibinfo{year}{1997}),
  \urlprefix\url{http://link.aps.org/doi/10.1103/PhysRevLett.78.2803}.

\bibitem[{\citenamefont{Neuenhahn and Marquardt}(2012)}]{Neuenhahn2012}
\bibinfo{author}{\bibfnamefont{C.}~\bibnamefont{Neuenhahn}} \bibnamefont{and}
  \bibinfo{author}{\bibfnamefont{F.}~\bibnamefont{Marquardt}},
  \bibinfo{journal}{Phys. Rev. E} \textbf{\bibinfo{volume}{85}},
  \bibinfo{pages}{060101} (\bibinfo{year}{2012}),
  \urlprefix\url{http://link.aps.org/doi/10.1103/PhysRevE.85.060101}.

\bibitem[{\citenamefont{Ros et~al.}(2015)\citenamefont{Ros, Mueller, and
  Scardicchio}}]{Ros2015}
\bibinfo{author}{\bibfnamefont{V.}~\bibnamefont{Ros}},
  \bibinfo{author}{\bibfnamefont{M.}~\bibnamefont{Mueller}}, \bibnamefont{and}
  \bibinfo{author}{\bibfnamefont{A.}~\bibnamefont{Scardicchio}},
  \bibinfo{journal}{Nuclear Physics B} \textbf{\bibinfo{volume}{891}},
  \bibinfo{pages}{420 } (\bibinfo{year}{2015}), ISSN \bibinfo{issn}{0550-3213},
  \urlprefix\url{http://www.sciencedirect.com/science/article/pii/S0550321314003836}.

\bibitem[{\citenamefont{Rigol}(2009{\natexlab{b}})}]{Rigol2009}
\bibinfo{author}{\bibfnamefont{M.}~\bibnamefont{Rigol}},
  \bibinfo{journal}{Phys. Rev. Lett.} \textbf{\bibinfo{volume}{103}},
  \bibinfo{pages}{100403} (\bibinfo{year}{2009}{\natexlab{b}}),
  \urlprefix\url{http://link.aps.org/doi/10.1103/PhysRevLett.103.100403}.

\bibitem[{\citenamefont{{Popescu} et~al.}(2005)\citenamefont{{Popescu},
  {Short}, and {Winter}}}]{Popescu2005}
\bibinfo{author}{\bibfnamefont{S.}~\bibnamefont{{Popescu}}},
  \bibinfo{author}{\bibfnamefont{A.~J.} \bibnamefont{{Short}}},
  \bibnamefont{and} \bibinfo{author}{\bibfnamefont{A.}~\bibnamefont{{Winter}}},
  \bibinfo{journal}{eprint arXiv:quant-ph/0511225}  (\bibinfo{year}{2005}),
  \eprint{quant-ph/0511225}.

\bibitem[{\citenamefont{Popescu et~al.}(2006)\citenamefont{Popescu, Short, and
  Winter}}]{Popescu2006}
\bibinfo{author}{\bibfnamefont{S.}~\bibnamefont{Popescu}},
  \bibinfo{author}{\bibfnamefont{A.~J.} \bibnamefont{Short}}, \bibnamefont{and}
  \bibinfo{author}{\bibfnamefont{A.}~\bibnamefont{Winter}},
  \bibinfo{journal}{Nat Phys} \textbf{\bibinfo{volume}{2}},
  \bibinfo{pages}{754} (\bibinfo{year}{2006}), ISSN \bibinfo{issn}{1745-2473},
  \urlprefix\url{http://dx.doi.org/10.1038/nphys444}.

\bibitem[{\citenamefont{Linden et~al.}(2009)\citenamefont{Linden, Popescu,
  Short, and Winter}}]{Linden2009}
\bibinfo{author}{\bibfnamefont{N.}~\bibnamefont{Linden}},
  \bibinfo{author}{\bibfnamefont{S.}~\bibnamefont{Popescu}},
  \bibinfo{author}{\bibfnamefont{A.~J.} \bibnamefont{Short}}, \bibnamefont{and}
  \bibinfo{author}{\bibfnamefont{A.}~\bibnamefont{Winter}},
  \bibinfo{journal}{Phys. Rev. E} \textbf{\bibinfo{volume}{79}},
  \bibinfo{pages}{061103} (\bibinfo{year}{2009}),
  \urlprefix\url{http://link.aps.org/doi/10.1103/PhysRevE.79.061103}.

\bibitem[{\citenamefont{Yukalov}(2011{\natexlab{a}})}]{Yukalov2011}
\bibinfo{author}{\bibfnamefont{V.}~\bibnamefont{Yukalov}},
  \bibinfo{journal}{Laser Physics Letters} \textbf{\bibinfo{volume}{8}},
  \bibinfo{pages}{485} (\bibinfo{year}{2011}{\natexlab{a}}), ISSN
  \bibinfo{issn}{1612-202X},
  \urlprefix\url{http://dx.doi.org/10.1002/lapl.201110002}.

\bibitem[{\citenamefont{Yukalov}(2012)}]{Yukalov2012}
\bibinfo{author}{\bibfnamefont{V.}~\bibnamefont{Yukalov}},
  \bibinfo{journal}{Physics Letters A} \textbf{\bibinfo{volume}{376}},
  \bibinfo{pages}{550 } (\bibinfo{year}{2012}), ISSN \bibinfo{issn}{0375-9601},
  \urlprefix\url{http://www.sciencedirect.com/science/article/pii/S0375960111013685}.

\bibitem[{\citenamefont{Yukalov}(2011{\natexlab{b}})}]{Yukalov2011a}
\bibinfo{author}{\bibfnamefont{V.}~\bibnamefont{Yukalov}},
  \bibinfo{journal}{Physics Letters A} \textbf{\bibinfo{volume}{375}},
  \bibinfo{pages}{2797 } (\bibinfo{year}{2011}{\natexlab{b}}), ISSN
  \bibinfo{issn}{0375-9601},
  \urlprefix\url{http://www.sciencedirect.com/science/article/pii/S0375960111007353}.

\end{thebibliography}

\end{document}